\documentclass{ws-mpla}

\usepackage{amssymb,amsbsy,amsmath,amsfonts}
\usepackage{graphicx}% Include figure files
\usepackage{float}
\usepackage{rotating}

\begin{document}

\markboth{L. S. GENG, J. MARTIN CAMALICH, L. ALVAREZ-RUSO, and M. J. VICENTE VACAS}
{ChPT study of the $N\rightarrow\Delta$ axial transition FF}

%%%%%%%%%%%%%%%%%%%%% Publisher's Area please ignore %%%%%%%%%%%%%%
%\catchline{}{}{}{}{}
%%%%%%%%%%%%%%%%%%%%%%%%%%%%%%%%%%%%%%%%%%%%%%%%%%%%%%%%%%%%%%%%%%%

\title{Chiral perturbation theory study of the axial $N\rightarrow\Delta(1232)$ transition}

\author{L. S. GENG, J. MARTIN CAMALICH, L. ALVAREZ-RUSO, and M. J. VICENTE VACAS}

\address{Departamento de F\'{\i}sica Te\'orica and IFIC, Centro
Mixto, Institutos de Investigaci\'on de Paterna - Universidad de
Valencia-CSIC}

\maketitle

\pub{Received (Day Month Year)}{Revised (Day Month Year)}

\begin{abstract}
We have performed a theoretical study of  the axial Nucleon to Delta(1232) ($N\rightarrow\Delta$)
transition form factors up to one-loop order in 
covariant baryon chiral perturbation theory within a
formalism in which the unphysical
spin-1/2 components of the $\Delta$ fields are decoupled.
\end{abstract}

\ccode{PACS Nos.: 23.40.Bw,12.39.Fe, 14.20.Gk.}

\section{Introduction}	
Nowadays, it is generally accepted that Quantum Chromodynamics (QCD) is the 
theory of the strong interaction. It has been very successful and tested to great precision 
at high energies; however,
its application in the low energy region of $\sim$ 1 GeV is quite problematic due to the large running coupling constant. The advent of
chiral perturbation theory ($\chi$PT)
 and lattice QCD approach has made possible a model independent study of the low-energy strong phenomena for the first time.

Neutrino physics has made remarkable progress in recent years,
as evidenced by the 2002 Nobel prize in physics (awarded partly to Raymond Davis Jr and
Masatoshi Koshiba ``for pioneering contributions to astrophysics, in particular for the detection of cosmic neutrinos''). After many years of experimental (and theoretical) efforts, two facts have been firmly established: (i) neutrino have
masses and (ii) different flavors of neutrino can oscillate into each other.
Presently, one of the main goals in the field is to measure accurately the masses and oscillation parameters. 
A good understanding of pion production is relevant to reduce systematic uncertainties
in oscillation experiments.
The axial nucleon to $\Delta$(1232) transition, characterized by four
form factors, plays an important role in this reaction at low $Q^2$ transfer.\cite{AlvarezRuso:2007it}

Most of our current (experimental) knowledge of the $N\rightarrow\Delta$ axial 
transition form factors
comes from neutrino bubble chamber data.\cite{Kitagaki:1990vs}
The possibility to extract them 
using parity-violating electron scattering at Jefferson Lab has been extensively studied,\cite{Mukhopadhyay:1998mn} and could  shed new light
on the nature of these form factors.
Present and future neutrino experiments (MiniBoone, K2K, Fermilab) could also provide further information.

In the past, the
theoretical descriptions have been done using different
approaches, mostly quark models (for a review, see Ref.~\refcite{Liu:1995bu}). 
In recent years,
there has been an increasing interest on these form factors. They have
been calculated, for instance, using the
chiral constituent quark model and 
light cone QCD sum rules. State of the art
calculations within lattice QCD
have also become
available.\cite{Alexandrou:2006mc}

While the axial $N\rightarrow\Delta$ form factors have been addressed in (tree level) HB$\chi$PT,\cite{Zhu:2002kh} \textit{no calculation has been performed up to now within the relativistic framework}. With lattice QCD results becoming
available and in view of the many ongoing experimental efforts to
extract these form factors from electron- and neutrino-induced reactions, it is timely to study the axial $N\rightarrow\Delta$
transition form factors within covariant baryon $\chi$PT.\cite{Geng:2008bm}

\section{Theoretical framework: covariant baryon $\chi$PT with explicit $\Delta$'s}
The study of the $N\rightarrow\Delta$ transition form factors using covariant baryon
$\chi$PT is much more complicated than it seems to be. To begin with, one has to address
the following three questions: power counting, chiral Lagrangians, and the appropriate form of
the $\Delta$ propagator.
\begin{enumerate}
 \item A proper power counting scheme is at the center of effective field theories.
To include the $\Delta(1232)$ explicitly, one has to count the $N$-$\Delta$ mass
difference $\varDelta \equiv M_\Delta-M_N\sim0.3$\,GeV properly. In the present work,
we adopt the $\delta$ expansion scheme, which counts 
$m_\pi/\Lambda_{\chi\mathrm{SB}}$ as $\delta^2$ to maintain
the scale hierarchy $m_\pi\ll\varDelta\ll\Lambda_{\chi\mathrm{SB}}$.\cite{Pascalutsa:2003zk}

\item The pion-nucleon and pion-pion Lagrangians are rather standard.
The $N\Delta$ and $\Delta\Delta$
Lagrangians, on the other hand, require more attention. The $\Delta(1232)$ is a spin-3/2 resonance and, therefore, its spin
content can be described in terms of the Rarita-Schwinger (RS) field,
$\Delta_\mu$, where $\mu$ is the Lorentz index.
This field, however, contains unphysical spin-1/2 components. In order to tackle this problem,
 we follow Ref.~\refcite{Pascalutsa:2006up} and adopt the
``consistent'' couplings, which are gauge-invariant under the
transformation
\begin{equation}
\Delta_\mu(x)\rightarrow\Delta_\mu(x)+\partial_\mu\epsilon(x).
\end{equation}

\item Different forms of the spin-3/2 propagator have been used in the literature,
some of which may lead to serious theoretical problems.\cite{Benmerrouche:1989uc}
Due to the spin-3/2 gauge symmetric nature of
the consistent couplings, we can use the most general spin-3/2 free
field propagator.\cite{Pascalutsa:2005nd}

\end{enumerate}
A more detailed discussion of these issues and the relevant $N\Delta$ and
$\Delta\Delta$ Lagrangians can be found in Ref.~\refcite{Geng:2008bm}.

\section{Results and Discussions}
\begin{figure}[t]
\centering
\includegraphics[scale=0.5]{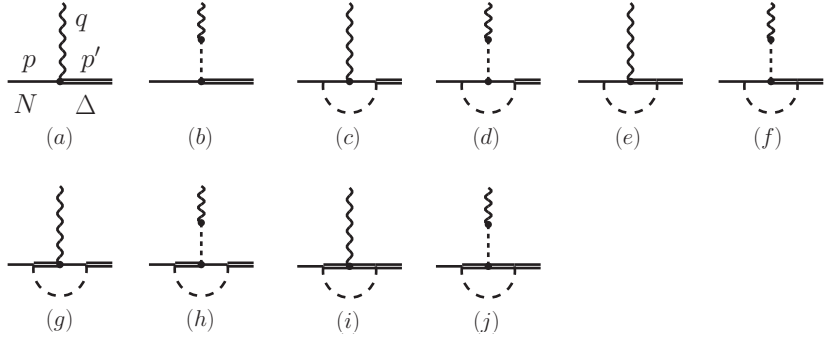}
\caption{Feynman diagrams contributing to the $N\rightarrow\Delta$
axial transition form factors up to order $\delta^{(3)}$. The double, solid, and
dashed lines correspond to the delta, nucleon, and pion, respectively; while the
wiggly line denotes the external pseudovector
source.\label{fig_diagram}}
\end{figure}

The $N\rightarrow\Delta$ axial transition form factors can be parametrized in
terms of the usually called Adler form factors:\cite{LlewellynSmith:1971zm,Schreiner:1973mj}
\begin{eqnarray*}\label{Adlerff}
\langle\Delta^+_{\alpha}(p')|-A^{\alpha\mu,3}|P(p)\rangle&=&
 \bar{\Delta}^+_{\alpha}(p')\bigg\{
 \frac{C^A_3(q^2)}{M_N}\big(g^{\alpha\mu}\gamma\cdot
 q-q^\alpha\gamma^\mu\big)\nonumber\\
&&\hspace{-1cm}+
 \frac{C^A_4(q^2)}{M^2_N}\big(q\cdot p' g^{\alpha\mu}- q^\alpha p'^\mu
\big)+C^A_5(q^2)g^{\alpha\mu}+\frac{C^A_6(q^2)}{M^2_N}q^\alpha
 q^\mu\bigg\}N,
\end{eqnarray*}
where $A^{\alpha\mu,3}$ is the third isospin component of the axial current.

In the $\delta$ expansion scheme, up to order $\delta^{(3)}$, all diagrams contributing to
the axial $N\rightarrow\Delta$ transition form factors are displayed in Fig.~\ref{fig_diagram}. The corresponding results in terms of low-energy constants (LEC) and
loop functions are summarized in Table \ref{table:avf}.

We can easily see that at $\delta^{(1)}$, $C^A_5(0)=\sqrt{\frac{2}{3}}\frac{h_A}{2}\approx1.16$, where
$h_A$ is the $\pi N\Delta$ coupling determined from the
$\Delta$ width, which is close to the Kitagaki-Adler value\cite{Kitagaki:1990vs} of 1.2.
The Kitagaki-Adlder assumption
$C^A_6=C^A_5\frac{M^2_N}{m^2_\pi-q^2}$, on the other hand, is satisfied only up to $\delta^{(2)}$, i.e., non pion-pole contributions appear at order $\delta^{(3)}$.\cite{Geng:2008bm}
\begin{table}[htpb]
     \tbl{\label{table:avf}The $N\rightarrow\Delta$ axial transition form factors in covariant baryon $\chi$PT;
 $d_1$, $d_2$, $d_3$, $d_4$ are order 2 LEC (in GeV$^{-1}$) while $f_1$, $f_2$, $f_3$, $f_4$, $f_5$, $f_6$, $f_7$ are order 3 LEC (in GeV$^{-2}$); $g_3(q^2)$, $g_4(q^2)$, $g_5(q^2)$, and $g_6(q^2)$ are the one-loop contributions as defined in Eq.~(31) of Ref.~7.}{\begin{tabular}{c|ccc}
     \hline\hline
      FF &  $\delta^{(1)}$ & $\delta^{(2)}$ & $\delta^{(3)}$\\
     \hline
$ -\sqrt{\frac{3}{2}}\frac{C^A_3(q^2)}{M_N}$& 0&$-d_2$ &$f_3\varDelta+g_3(q^2)$\\
 $-\sqrt{\frac{3}{2}}\frac{C^A_4(q^2)}{M^2_N}$&0 &$-d_1/M_\Delta$&$ (f_4+f_6)\varDelta/M_\Delta+g_4(q^2)$\\
$- \sqrt{\frac{3}{2}}C^A_5(q^2)$&$-\frac{h_A}{2}$&$-(d_3+d_4)\varDelta$&$(f_5+f_7) \varDelta^2+(f_1+f_2)q^2 +g_5(q^2)$\\
 $-\sqrt{\frac{3}{2}}\frac{C^A_6(q^2)}{M^2_N}$&$\frac{h_A/2}{q^2-m^2_\pi}$&$\frac{(d_3+d_4) \varDelta}{q^2-m^2_\pi}$&$-f_1+g_6(q^2)+\frac{-(f_5+f_7)\varDelta^2-f_2
 q^2-(g_5(q^2)+g_6(q^2) q^2)}{q^2-m^2_\pi}$\\
    \hline\hline
    \end{tabular}} % \par
       \end{table}

The form factors $C^A_3$ and $C^A_4$ both start at chiral order 2 and get their $q^2$ dependence at order 3 from the loops.
For $C^A_3$, we find a  small $q^2$ dependence, which is quite sensitive to the $\pi\Delta\Delta$ coupling constant. On the other hand, its imaginary part, coming mainly from
the $N$-$N$ internal diagram, is finite ($\sim0.03$ at $q^2=0$) and has a mild $q^2$ dependence. This suggest that $C^A_3$ is small (compared to $C^A_{4,5,6}$) but not necessarily zero. The $C^A_4$ dependence on $q^2$  is also found to be rather mild at order $\delta^{(3)}$. 

In covariant $\chi$PT up to $\delta^{(3)}$, four $\delta^{(2)}$ and seven $\delta^{(3)}$ LEC appear in the
results. However, some of them appear in particular combinations. Therefore, effectively we have only
five unknown constants. They can be fixed by fitting either to the phenomenological form factors obtained from neutrino bubble chamber data (with several assumptions), to the results of other approaches, such as
those of various quark models, or to the lattice QCD results\cite{Alexandrou:2006mc}. For a more detailed discussion, see Ref.~\refcite{Geng:2008bm}.

\section*{Acknowledgments}

We thank Mauro Napsuciale, Stefan Scherer, Wolfram Weise,
 and in particular Massimiliano Procura and Vladimir Pascalutsa for useful
discussions.  L. S. Geng acknowledges financial
support from the Ministerio de Educacion y Ciencia in the Program 
``Estancias de doctores y tecnologos extranjeros''. J. Martin Camalich acknowledges  the same institution for a FPU fellowship. This work was partially supported by the  MEC 
contract  FIS2006-03438, the Generalitat Valenciana ACOMP07/302,
and the EU Integrated Infrastructure
Initiative Hadron Physics Project contract RII3-CT-2004-506078.

 %\bibliographystyle{plain}
 %\bibliography{chiral}

\end{document}